\documentclass[pdftex,twocolumn,epjc3]{svjour3}    

\RequirePackage[T1]{fontenc}

\smartqed 
\usepackage{graphicx}
\usepackage{braket}
\usepackage[utf8]{inputenc}
\usepackage{amsmath}
\usepackage{amssymb}
\usepackage{bbold}
\usepackage{amsfonts,bm}
\usepackage{geometry}
\usepackage{fullpage}
\usepackage{caption}
\usepackage{todonotes}
\usepackage{multicol}
\usepackage{subcaption}
\usepackage{slashed}
\usepackage{float}
\usepackage{epstopdf}
\usepackage{units}
\usepackage{bm,paralist,xspace,url,relsize}
\usepackage{multirow}
\usepackage{isotope}
\RequirePackage{mathptmx}   
\RequirePackage{flushend}
\RequirePackage[numbers,sort&compress]{natbib}
\RequirePackage[colorlinks,citecolor=blue,urlcolor=blue,linkcolor=blue]{hyperref}
\graphicspath{{./Plots/}}
\usepackage[normalem]{ulem}

\definecolor{OliveGreen}{rgb}{0,0.6,0}
\definecolor{amber}{rgb}{1.0, 0.75, 0.0}

\newcommand{\la}{\langle}
\newcommand{\ra}{\rangle}
\newcommand{\Tm}{\isotope[171]{Tm}}
\newcommand{\Sm}{\isotope[151]{Sm}}

\journalname{Eur. Phys. J. A}

\begin{document}

\title{Can we use heavy nuclei to detect relic neutrinos?}

\author{Oleksii~Mikulenko\thanksref{e1,addr1}       
        \and
        Yevheniia~Cheipesh\thanksref{addr1}  \and
        Vadim~Cheianov\thanksref{addr1} \and
        Alexey~Boyarsky\thanksref{addr1} 
}

\thankstext{e1}{e-mail: mikulenko@lorentz.leidenuniv.nl}

\institute{Instituut-Lorentz, Leiden University, Niels Bohrweg 2, 2333 CA Leiden, The Netherlands\label{addr1}
}

\date{Received: date / Accepted: date}

\maketitle

\begin{abstract}
Recent analysis of the viability of solid state-based relic neutrino detectors has revealed the fundamental necessity for the use of heavy, $A>100$, $\beta$-decayers as neutrino targets. Of all heavy isotopes, $\Tm$ and $\Sm$ stand out for their sufficiently low decay energies, reasonable half-life times and stable daughter nuclei. However, the crucial bit of information, that is the soft neutrino capture cross-section is missing for both isotopes. The main reason for that is a particular type of $\beta$-decay, which precludes a simple link between the isotope's half-life time and the neutrino capture rate. In light of the necessity for a reliable estimate of the capture rate --- unimpeded by potentially devastating theoretical uncertainties --- prior to using the isotope in a full-scale experiment, we propose an experimental method to bypass this difficulty and obtain the capture cross-section of a soft neutrino by a given isotope from the isotope's $\beta$-spectrum.
\end{abstract}

\section{Introduction}
\label{section:intro}
The ambitious goal of detection~\cite{Weinberg:1962zza} and the measurement of the mass~\cite{Formaggio:2021nfz} of the relic neutrino relies on the precise experimental knowledge of the 
$\beta$-spectrum of radioactive elements~\cite{Weinheimer:2013hya, Betti:2019ouf}. Relic neutrinos, which fill the totality of space in the form of an almost ideal gas of temperature $T_{\nu} \approx \unit[1.95]{K},$ are expected to manifest themselves in rare neutrino capture events. 
Such events involving cosmic neutrinos of mass $m_\nu$ and a sample of radioactive atoms characterized by the $\beta$-decay energy $Q$
would produce an extremely faint peak at the energy $Q+m_\nu c^2$ in the $\beta$-spectrum of the sample.
We recall that for all radioactive elements the overwhelming bulk 
of the $\beta$-spectrum arises from spontaneous $\beta$-decay and forms a continuum with the upper cutoff energy $Q-m_{\nu}^0 c^2$ where 
$m_{\nu}^0$ is the mass of the lightest neutrino. For this reason one expects the neutrino capture peak to be separated from the end of the spontaneous $\beta$-spectrum by an energy gap of at least 
one neutrino mass and for that reason to be discernible at least 
in principle.

Despite the simplicity of its theoretical premise, a neutrino capture experiment establishing the existence of relic neutrinos 
has not yet materialized. The reason for this is the  weakness  of  the neutrino-matter interaction,  which  makes  it  difficult  to achieve the sufficient  number of capture events in a reasonably sized radioactive sample. The requirement of a large neutrino 
capture cross-section combined with other important 
considerations such as the manageable half-life time and 
the stability of the daughter isotope turn out to be 
so restrictive that only a handful of atoms can be viewed
as viable candidates for the $\rm C\nu B$ detection 
experiment. From this perspective, Tritium has 
long been regarded as the best candidate $\beta$-emitter~\cite{Betti:2019ouf, Wolf:2008hf, Cocco:2007za, Lazauskas:2007da, Blennow:2008fh, Li:2010sn, Faessler:2011qj, Long:2014zva}, even though 
it was found that the workable sample of gaseous molecular 
Tritium falls short of the required activity levels by six orders of magnitude. Currently, the only viable alternative to the gas phase experiment is a solid state based architecture where the atomic tritium is adsorbed on a substrate~\cite{Betti:2019ouf}.

The low event rate is not the only hindrance 
in the way of relic neutrino detection. The upper bounds on the neutrino mass~\cite{Loureiro:2018pdz} show that the energy gap between the signal from neutrino capture and the background is extremely small $m_\nu /Q \ll 1$, therefore the detection
of the $\rm C\nu B$ requires extraordinary energy resolution. It has been demonstrated that the electromagnetic guidance 
system and the calorimetry module of the detection apparatus can 
be built to such stringent specifications~\cite{Betti:2019ouf}. However, solid-state interactions introduce additional complications~\cite{Cheipesh:2021fmg, Tan:2022eke, Nussinov:2021zrj, PTOLEMY:2022ldz, Catena:2023qkj}. In particular, it has been shown in~\cite{Cheipesh:2021fmg} that deposition of $\beta$-emitters on a solid-state substrate 
produces a new fundamental limitation on the experimental resolution originating in the zero-point motion of the 
emitter's centre of mass. For Tritium on solid surfaces, 
the best theoretical resolution is $\Delta E \sim \unit[0.5]{eV}$ which is an order of magnitude worse than what is required in order to see the relic neutrino peak. Furthermore, it was shown~\cite{Cheipesh:2021fmg} that the main factor that determines this value is the ratio of the $\beta$-decay energy $Q$ to the mass of the emitter nucleus $m_{\text{nucl}}$, namely 
\mbox{$\gamma =[Q^2 m_{\text{e}}/
m_{\text{nucl}}^3]^{1/4}.$} 
This finding opens a new avenue to search for a possible alternative for Tritium that would have both a sufficient event rate and low enough energy uncertainty. In the same work~\cite{Cheipesh:2021fmg}, it was found that the two promising candidates that have low enough $\gamma$-values are Thulium ($\Tm$) and Samarium ($\Sm$) with $\gamma_{\isotope[3]{H}}/\gamma_{\Tm} = 0.11$ and  $\gamma_{\isotope[3]{H}} / \gamma_{\Sm} = 0.1$ respectively. This means that the intrinsic energy uncertainty for these isotopes is an order of magnitude smaller than that of Tritium. This value approaches the upper bound for the neutrino mass  and therefore could, in principle, provide sufficient energy resolution for its detection.

The $\gamma$-value introduced in the previous paragraph is 
defined in terms of the simple intrinsic characteristics of a nucleus such as its mass and $Q$-value and therefore is straightforward to calculate. 
In contrast, the neutrino capture cross-section has not been calculated for every isotope. In particular, it is not known 
for either of the isotopes of interest, $\Tm$ and $\Sm$.

The reason for this emanates from the complexity of the nuclear structure~\cite{siegbahn1955beta, BookKonopinski, BookSchopper}. The $n\leftrightarrow p$ conversion changes the nuclear state, introducing some unknown amplitudes of transition between the initial and final nuclear wave-functions. For some isotopes, there is a single dominant transition. This allows to extract the corresponding matrix element solely on the basis of the nucleus lifetime $\tau$ of weak decay. Other isotopes undergo the so-called \textit{non-unique forbidden transition}, which, as the name suggests, involves multiple possible configurations of the final nucleus with independent transition amplitudes. In this case, the direct link between the observed half-life time and the neutrino capture rate is generally absent.

A possible way to confront this difficulty relies on performing theoretical computations of the nuclear matrix elements. The list of nuclear models includes but is not limited to Nuclear Shell Model~\cite{RevModPhys.77.427} and Interacting Boson-Fermion model \cite{Arima:1981hp, Iachello:2005aqa}. These have been employed in the work~\cite{Kostensalo:2022tcs} to evaluate the relic neutrino capture cross-section for isotopes $\Tm$, $\Sm$, and $\isotope[210]{Pb}$ after the completion of this work.

The goal of the present paper is to show how the neutrino capture cross section can be extracted from the \textit{experimentally accessible} $\beta$-spectrum for a given radioactive isotope decaying through non-unique forbidden transitions.

\section{Quantum mechanics of $\beta$-interaction and crude estimate of neutrino capture}\label{section:QM}

Neutrino capture and $\beta$-decay are the same processes driven by the weak interaction; they differ only in whether the (anti)neutrino is in the initial or final state. To establish the exact connection between their respective rates, we start from briefly reminding the main concepts of $\beta$-decay theory. We consider the sibling processes of $\beta$-decay and neutrino capture by a generic nucleus
\begin{align}\label{eq:decays}
   \nonumber (A,Z) &\rightarrow (A, Z+1) + e^{-} + \bar{\nu}_
    e \\
    \nu_e + (A,Z) &\rightarrow (A, Z+1) + e^{-}.
\end{align}
which are driven by the same weak $\beta$-decay Hamiltonian
\begin{equation}\label{eq:Hamiltonian}
    \mathcal{H}^\beta = \dfrac{G_\beta}{\sqrt{2}}\bar{\psi}_e\gamma^\mu (1-\gamma_5)\psi_\nu\,\,\bar{p} \gamma_\mu (g_V + g_A\gamma_5)n +\text{h.c.},
\end{equation}
where $G_\beta = G_F\cos\theta_C$ and $\theta_C$ is Cabbibo angle, $\psi_e$, $\psi_\nu$ are electron and neutrino fields and $p$, $n$ being the proton and neutron fields respectively. The vector $g_V$ and axial $g_A$ coupling constants are renormalized by strong interactions with $|g_A/g_V|\approx 1.27$~\cite{Gell-Mann:1960mvl, ParticleDataGroup:2020ssz}.

The differential $\beta$-decay rate $d\Gamma_\beta$ and the capture cross-section for spin-averaged neutrino are given by the Fermi 
Golden Rule and can be written as\footnote{Here we use the fact that absorption of antineutrino with momentum $p_\nu$ is equivalent to emission of neutrino with momentum $-p_\nu$}:
\begin{align}\label{eq:both_rates}
    \nonumber d\Gamma_\beta &= \dfrac{1}{2\pi^3}\times  p_\nu E_\nu p_eE_e dE_e \times W_\beta(p_e, p_\nu) \\
    (\sigma v)_\nu &= \lim_{p_\nu\to 0} \dfrac{1}{\pi}\times p_eE_e \times W_\nu (p_e, p_\nu) ,
\end{align}
where $p_{e(\nu)}$ and $E_{e(\nu)}$ are the 
momenta and energies of the leptons, $W_\beta(p_e, p_\nu)$ is 
the average transition rate for the decay 
of an atom into two lepton plane waves with momenta 
$p_e, p_\nu,$ and $W_\nu(p_e,p_\nu)$ is the average 
transition rate for the capture of a 
neutrino having the momentum $p_\nu$ and the 
emission of an electron with momentum $p_e.$

The average transition rates are expressed in terms of
transition amplitudes by
\begin{equation}
    W_{\beta, \nu }(p_e, p_\nu) = \int \frac{d\Omega_e}{4\pi} \frac{d\Omega_\nu}{4\pi} \sum \vert \mathcal M_{\text{if}}^{\beta, \nu }(\mathbf p_e, \mathbf p_\nu)\vert^2.
    \label{eq:Wdef}
\end{equation}
Here $\mathcal M_{\text{if}}$ is the quantum transition 
amplitude between the initial and the final state
induced by the reduced weak interaction 
Hamiltonian~\cite{BookSchopper,BookKonopinski,buhring1963beta}
\begin{equation}
\label{eq:matrix_element}
    \mathcal M_{\text{if}} = \frac{G_\beta}{\sqrt{2}} \int \bar \psi_e (\mathbf r) \gamma^\mu (1-\gamma_5) \psi_\nu(\mathbf r)  J^\mu_{\text{nuclear}}(\bm r) \,d\bm r ,
\end{equation}
which encapsulates all information about the changes
in the internal nuclear structure in a function $J^\mu_{\text{nuclear}}(\mathbf{r}).$  This function 
cannot be calculated from first principles. However, its transformation properties under the symmetry group of 
space are known for each transition. The summation symbol 
in Eq.~\eqref{eq:Wdef} is a shorthand for the sum 
over the spin quantum numbers of the out-states 
as well as averaging over the spins of the 
in-states. The averaging over the directions of 
$\mathbf p_e$ and $\mathbf p_\nu$ is shown 
explicitly. Two important remarks are in order 
\begin{itemize}
    \item[1] For an overwhelming part of the $\beta$-spectrum 
    one can consider the neutrino as a massless (Weyl) 
    particle in both 
    the energy conservation law and the wave functions entering the transition amplitudes. There exists a tiny energy window 
    on the order of $m_\nu$ near the high-energy end of the 
    $\beta$-spectrum where the neutrino mass plays a role. However, the resolution required for the observation of 
    the $\beta$-spectrum inside that window is by far beyond the reach of the  existing experimental technique. Since the 
    existing $\beta$-decay experiment cannot distinguish between 
    the massive and massless cases, {\em we shall throughout this note discuss the function $W_\beta(p_e, p_\nu)$ 
    assuming the $m_\nu\to 0 $ limit. }
    \item[2] Our main focus is on neutrino capture processes 
    involving the cosmic neutrino background. Relic neutrinos are non-relativistic 
    $p_\nu \ll m_\nu,$ which is the opposite of the ultra-relativistic limit discussed in item 1. It is straightfoward to see that: 
    \begin{equation}
        W_\nu(p_e, 0) = \frac{1}{2} \lim_{p_\nu \to 0} 
        W_\beta (p_e, p_\nu)
        \label{eq:onehalf}
    \end{equation}
    for a left-handed particle with a Majorana mass term.
    Indeed, in the $p_\nu\to 0$ limit the incoming massive neutrino is a superposition of a left-handed Weyl particle and a right-handed Weyl anti-particle $$\vert \text{Majorana} \rangle = (\vert \nu \rangle + \vert \bar \nu  \rangle)/\sqrt 2. $$ In a process where an electron is created, the operator \eqref{eq:matrix_element} only picks one term of the two, hence 
    the corresponding transition rate is one half of the transition 
    rate $W_\beta$ of a Weyl neutrino.
\end{itemize}

\subsection{Crude estimate of neutrino capture}
In this subsection, we want to provide a simple order-of-magnitude estimate for neutrino capture cross-section. To this end, we assume that the matrix element has no dependence on the lepton energy and reduces to a constant encoding the information about the initial and final nuclear states
\begin{equation}
\label{eq:expansion0}
    \sum |\mathcal M^\beta_{\text{if}}(p_e, p_\nu)|^2 = \text{const}. \end{equation}
Such an approximation neglects the Coulomb interaction between the emitted electron 
and the nucleus. 

In this case, all the structural information about the nuclei gets absorbed into a constant numerical factor, therefore the ratio of the $\beta$ decay and the neutrino capture rates, Eqns.~\eqref{eq:both_rates}, is 
completely determined by the phase volume factors $p_\nu^2p_eE_e$ and $p_e E_e$ accordingly. Using Eq.~\eqref{eq:onehalf}, this gives rise to the following 
relationship between the capture cross-section $\left(\sigma v\right)_\nu,$  the total lifetime $\tau = (\int d\Gamma_\beta)^{-1}$ of a $\beta$-decaying isotope, and the total kinetic energy $Q$ released in the reaction:
\begin{equation}
\label{eq:preestimate}
    (\sigma v)_\nu = \tau^{-1} \frac{(2\pi)^{-1} p_e E_e}{(2\pi^3)^{-1}\int^{m_e + Q}_{m_e} E_e' p_e' (Q-T_e')^2 \, dE_e'},
\end{equation}
with $T_e = E_e-m_e$ being the kinetic energy of the electron, and neutrino momentum in $\beta$ decay is $p_\nu = Q-T_e$. 
In the particular case of nonrelativistic electron $Q\ll m_e$, this relation gives the following simple scaling:
\begin{equation}
\label{eq:simple_sigma}
    (\sigma v)_{\text{est.}} = \unit[5.3\cdot 10^{-46}]{cm^2} \times \frac{\unit[1]{year}}{\tau} \times \left(\frac{\unit[100]{keV}}{Q}\right)^3.
\end{equation}

In order to quantify the error introduced by the simplifying assumptions leading up to Eq.~\eqref{eq:expansion0},
we introduce a correction factor $\delta$ such that the actual cross-section is given by
\begin{equation}
\label{eq:delta}
    (\sigma v)_\nu  = \delta \times (\sigma v)_{\text{est.}}
\end{equation}
The values of $\delta$ for a number of elements where the exact results for the neutrino 
capture cross-section are known~\cite{Cocco:2007za} are given in Tab.~\ref{tab:delta}. One can 
see that in all those cases $\delta$ is reasonably close to unity.

 \begin{table}[h!]
    \centering
    \begin{tabular}{|c|c|c|c|c|}
    \hline
         Isotope& $Q$, $\unit{keV}$& $\tau$, year & $(\sigma v)_\nu$, $\unit[10^{-46}]{cm^2}$ &$\delta$  \\
    \hline
        $\isotope[3]{H}$& $18.591$& $17.8$ & $39.2$ &$0.86$
        \\
    \hline
        $\isotope[63]{Ni}$& $66.945$& $145$ &  $6.9\cdot 10^{-2}$ &$0.57$
        \\
    \hline
        $\isotope[93]{Zr}$& $60.63$& $2.27\cdot 10^{6}$ &  $1.20\cdot  10^{-5}$ &$1.15$  
        \\
  \hline
        $\isotope[106]{Ru}$& $39.4$& $1.48$ &  $29.4$ &$0.51$ 
        \\
  \hline
        $\isotope[107]{Pd}$& $33$& $9.38\cdot 10^{6}$ &  $1.29\cdot 10^{-5}$ &$0.83$
        \\
 \hline
        $\isotope[187]{Re}$& $2.646$& $6.28\cdot 10^{10}$ &  $2.16\cdot 10^{-6}$ &$0.48$
        \\ 
        \hline
  \hline
     $\isotope[171]{Tm}$  & $96.5$ &  $1.92$  &  $2.1 \times  \delta$  & ---
        \\ 
  \hline
    $\isotope[151]{Sm}$ & $76.6$ & 90 & $0.091\times \delta$ & ---
        \\
    \hline
    \end{tabular}
    \caption{Neutrino capture cross-sections for different isotopes from~\cite{Cocco:2007za}. Note that $(\sigma v)_\nu$ differ from those of~\cite{Cocco:2007za} by a factor two due to neutrino spin averaging, as pointed out in~\cite{Long:2014zva}. One can see that the parameter $\delta$ defined by Eq.~\eqref{eq:delta} varies only by a factor of two from the identity that signals that Eq.~\eqref{eq:simple_sigma} gives a good approximation for the capture rates of the given isotopes.}\label{tab:delta}
    \end{table}

We are interested in neutrino capture by possible candidates for solid-state based C$\nu$B detection experiments --- $\Tm$ and $\Sm$. For these isotopes, the parameterization \eqref{eq:delta} reads
\begin{align}
    \nonumber (\sigma v)_{\Tm} & = \unit[2.1 \cdot 10^{-46}]{cm^{2}} \times \delta_{\Tm} \nonumber\\
    &\approx 0.054\,\, (\sigma v)_{\isotope[3]{H}}\times \delta_{\Tm} \\
     (\sigma v)_{\Sm} &=  \unit[9.1 \cdot 10^{-48}]{cm^{2}} \times \delta_{\Sm}\nonumber \\ &\approx 0.0023\,\, (\sigma v)_{\isotope[3]{H}} \times \delta_{\Sm}.
     \label{eq:deltafactors}
\end{align} 
However, unlike the isotopes listed in Table~\ref{tab:delta}, 
the theoretical values of the $\delta$ factors for $\Tm$ and 
$\Sm$ are not known. This is because both isotopes 
have a rather peculiar structure of the matrix element~\eqref{eq:matrix_element}, as explained in the following 
paragraph. 

We follow the general formalism presented in~\cite{Behrens:1971rbq, ElectronWaveFunctions}, see also \cite{Mougeot:2015bva} for a modern review. For purely illustrative purposes, we neglect the effect of the Coulomb attraction between the $\beta$-electron and the daughter nucleus, bearing in mind that in practice such an approximation may result in significant inaccuracy.  We recall that the function $J^\mu_{\text{nuclear}}(\bm r)$ is mainly localized inside the nucleus $r<R,$ and decays rapidly with increasing $r$ for $r>R$, where $R=A^{1/3} \times \unit[1.2]\times 10^{-13} \ \text{cm}$ is the radius of the nucleus. Since the typical lepton momentum is of the order $\unit[1]{MeV}\ll R^{-1}$, one can expand the matrix elements and the sum $\sum |\mathcal M_{\text{if}}|^2$ as 
a series in small parameters $p_{e/\nu} R \ll 1$\footnote{If Coulomb attraction is taken into account, the constants in this expansion get multiplied by correction factors $F_i(p_e)$, which do not depend on unknown nuclear physics and can be computed explicitly. }
\begin{equation}
\label{eq:expansion1}
    \sum |\mathcal M_{\text{if}}|^2 = c_{0} + c_1 \cdot p_e R + c_2 \cdot p_\nu R + \dots
\end{equation}
The constants $c_i$ in this expression are in essence 
combinations of the spherical multipole moments of 
$J^\mu_{\text{nuclear}}(\mathbf r)$ containing 
structural information about the many-body wave functions 
of the parent and daughter nuclei. The simplifying 
approximation~\eqref{eq:expansion0} amounts to 
keeping only the leading-order term $c_0$ in 
the expansion~\eqref{eq:expansion1}, which in many 
cases is well justified. For some isotopes, however, 
electroweak selection rules demand that $c_0=0$.
Indeed, if the mother and daughter isotopes have different 
spin and parity, at least one 
of the leptons is required to carry a non-vanishing 
orbital angular momentum. Since a lepton's wave function
corresponding to the orbital angular momentum $l$ 
has the asymptotic form $(p r)^l$ at small $r,$
the matrix element of such a transition, Eq.~\eqref{eq:matrix_element},
will necessarily contain terms proportional to 
$\left(p_{e} R\right)^l(p_\nu R)^{l'}$ with $l+l'>0.$ 
The worst case scenario, 
known as a {\it forbidden non-unique transition}, is 
when the selection rules admit for the presence 
of several commensurate leading-order terms on 
the right hand side of the asymptotic expansion 
Eq.~\eqref{eq:expansion1}. For such a transition the matrix 
element~\eqref{eq:expansion1} contains several 
unknown constants $c_i$, each multiplying 
its own unique function of energy. If that happens, 
the cancellation of the unknown constants, such as 
the one seen in Eq.~\eqref{eq:preestimate}, does not occur 
and the neutrino capture cross-section cannot 
be inferred from the isotope's life time. 

This conclusion may be relieved for heavy isotopes by the Coulomb attraction, which introduces two modifications. Firstly, the electronic wave-function are distorted by the presence of a ``point charge'', resulting in additional known energy-dependent coefficients in the expansion \eqref{eq:expansion1}. Secondly, the account for the non-zero size of the charge adds another dimensionless parameter $\xi \cdot R \equiv \alpha Z/R \cdot R$ to $p_{e,\nu} R$  in \eqref{eq:expansion1}. For heavy nuclei one can expect this parameter to be dominant if $\xi/Q\gg 1$. This motivates the so-called $\xi$-approximation, in which one keeps only expansion in the parameter $\alpha Z$ and is left with a single nuclear constant. In this case, the spectrum of isotopes that undergo non-unique transitions has the allowed shape, and the usual technique of relating the neutrino capture cross-section to the half-life time of the isotope may be used.

The theoretical motivation of the $\xi$-approximation does not guarantee that the associated nuclear constant is not suppressed. This can abate the dominance of the corresponding term and break down the approximation. Such a situation requires some degree of fine-tuning, similar to the case of isotopes whose spectra deviate from the expected allowed shape due to the suppression of the leading order term. While it may seem unlikely, it further reinforces the necessity of a method that can establish the neutrino capture cross-section in a more direct way.

Given the high stakes of the PTOLEMY experiment, we require a method to estimate the neutrino cross-section without relying on the assumption of no-cancellation. At the same time, for the purposes of this paper, we do not seek high accuracy of the result. It is important to note that, if the $\xi$-approximation is experimentally established to be accurate for a particular isotope, the relation of the cross-section to the half-life time may give much more precise results.

\section{Experimental determination of the neutrino capture rate 
from the end of the $\beta$ decay spectrum}

We have established that for isotopes such as $\Tm$ and $\Sm$
the knowledge of the lifetime and the $Q$-value is insufficient
in order to predict the neutrino capture cross-section. 
Here, we discuss how the required cross-section can be inferred directly from the experimentally measured $\beta$-spectrum. 
Our approach is based on two key observations. Firstly, 
both the emission and capture processes are governed by the 
same unknown structure function 
$W_\beta(p_e, p_\nu),$ albeit taken at different 
values of arguments. 
Specifically, a capture process corresponds to the limit
$p_\nu \to 0 $ and $$p_e = \sqrt{(Q+m_e)^2 - m_e^2},$$
whilst in a spontaneous $\beta$-decay process 
$$p_e=\sqrt{(Q+m_e-p_\nu)^2-m_e^2},$$ where 
$p_\nu$ can take any value between $0$ and $Q,$ resulting 
in a broad $\beta$-spectrum. Secondly, the function 
$W_\beta(p_e, p_\nu)$ is an analytic function of both arguments 
near the end point $p_\nu=0$ of the $\beta$-spectrum. 
We recall 
that in our discussion $W_\beta(p_e, p_\nu)$ is the rate involving transitions with massless neutrino states (see discussion at the end of section \ref{section:QM}).

Using the analyticity of $W_\beta(p_e, p_\nu)$ and making use of equations \eqref{eq:both_rates} and \eqref{eq:onehalf} we 
write the following expansion\footnote{Such a linear behaviour can be seen in the spectra generated by the BetaShape software, which predicts $(\sigma v)_\nu = \unit[1.2 \cdot 10^{-46}]{cm^{2}} (\Tm)$, $ \unit[4.8 \cdot 10^{-48}]{cm^{2}} (\Sm)$ and $\alpha_1 =0.25 (\Tm)$, $0.21 (\Sm)$.
For further discussion see Sec.~\ref{sec:conclusion}.}  for the observable $\beta$-spectrum near the edge $p_\nu = 0$
\begin{equation}
    \frac{\pi^2}{p_\nu^2 } \frac{d \Gamma_\beta}{dE_e}=
    (\sigma v)_\nu\times \left[1 + \alpha_1 p_\nu/Q + O(p_\nu^2/Q^2)\right]
    \label{eq:linear}
\end{equation}
where $\alpha_1$ is a constant. The characteristic energy 
scale where the linear approximation is 
applicable can be estimated from the microscopic theory 
of $\beta$ decay.
For the purposes of the present work, we notice that 
the physics of $\beta$ decay of heavy nuclei 
involves three important energy scales,
that are $Q,$ $m_e,$ and $1/R$. The smallest 
of the three defines the energy range where the 
expansion \eqref{eq:linear} works well. 
For $\Tm$ and $\Sm$ the smallest energy scale is $Q.$

Now we are in position to discuss the experimental procedure.
We assume a finite energy resolution $\Delta E $ of  the experiment (say, $\unit[1]{keV}$). We propose a way to deduce the neutrino capture rate of the $\Tm$ and $\Sm$ from the end of their experimentally measured $\beta$ spectra:
\begin{enumerate}
    \item Define some experimentally accessible energy resolution $\Delta E \ll Q$ and measure the number of $\beta$ decay events $N$ in several energy bins\footnote{We note that the spectrum itself behaves as $d\Gamma/dE_e \sim p^2_\nu$ and, therefore, events within a single bin are not uniformly distributed. Most of the events occur near the left side of a bin, which may introduce an additional systematic uncertainty. A possible way to avoid this problem and is to measure the integral number of events $N(p_\nu) = \int^{Q}_{Q-p_\nu} \frac{dN}{dT_e} dT_e$ and consider the function $N(p_\nu) \cdot p_\nu^{-3}$. This can be also fitted by a linear function and therefore used to extract $(\sigma v)_\nu$. In addition, this method allows to collect more statistics compared to the one with bins for sufficiently large $p_\nu$. } $T_e \in [Q-(n+1)\Delta E, Q-n\Delta E]$ as a function of the electron energy residue $\varepsilon_n = \Delta E(n+1/2)$ 
    \item We assume that all the decay events are detected. In this case, one can check whether the experimental points $N(\varepsilon_n)\times  (\unit[\varepsilon_n]{ in\,\,  keV})^{-2}$ fit the linear curve. If so, continue the obtained fit up till the value $\varepsilon_n = 0$.
    \item Assuming that the time of the measurement is $T_{\text{m}}\ll \tau$ and there are $N_{\text{at}}$ decaying atoms, the neutrino capture rate can be estimated as 
    \begin{multline}
    \label{eq:final_estimate}
       \left(\sigma v\right)_\nu =  \frac{\unit[7.0 \cdot 10^{-37}]{cm^2}}{(T_{\text{m}} \text{ in hours})(\Delta E \text{ in keV})}
       \\
       \times \frac{1}{N_{\text{at}}}\left(\dfrac{N(\varepsilon_n)}{(\unit[\varepsilon_n]{ in\,\,  keV})^2}\right)\bigg|_{\varepsilon_n =0}
    \end{multline}
\end{enumerate}

A remark should be made concerning the generality of~\eqref{eq:final_estimate}. Until now we neglected possible contributions to the electron spectrum due to $\beta$-decay into excited states of daughter nuclear or/and electronic shell of the atom. Let us comment on these contributions:
\begin{enumerate}
    \item Excited nuclear states have typical energies $E_{\text{ex}} \sim \unit[10]{keV}$, for instance, $\unit[66.7]{keV}$ for $\isotope[171]{Yb}$~\cite{Baglin:2018vrm} (daughter isotope for $\isotope[171]{Tm}$) and $\unit[21.5]{keV}$ of $\isotope[151]{Eu}$~\cite{Singh:2009ecb} (daughter isotope of $\isotope[151]{Sm}$). They do not contribute to the spectrum near the endpoint for $T_e >Q-E_\text{ex}$. Therefore, they are not relevant for the measurement of the spectrum high-energy tail with resolution of order 1 keV.
    \item Atomic excitations start to contribute to the spectrum from $\sim \unit[1]{eV}$ and are expected to be too small to be resolved with a typical energy resolution of an experiment. If this is the case, Eq.~\eqref{eq:final_estimate} includes all these transitions and overestimates the actual cross-section. For $Z\sim 60$, the probability to excite the electronic configuration is expected to be less than $30\%$~\cite{Carlson:1968zz, Hayen:2017pwg}, which translates into the same possible error in the value of the cross-section. The account for this effect may be done with the use of the atomic mismatch correction~\cite{Carlson:1970uip, Desclaux:1973qgc}. 
\end{enumerate}
The corrections discussed above may only introduce a difference by a prefactor of order one and therefore are beyond our considerations.

 \section{Conclusion and discussion}
 \label{sec:conclusion}

The most promising route towards the relic neutrino detection is currently through the use of solid state based 
detectors where the $\beta$ emitters are adsorbed on a substrate. 
Such a design has the potential to achieve sufficient 
density of emitters in a controllable way (such that electron scattering remains suppressed), and hence get a sufficient number of capture events. However, any $\beta$ decay experiment that uses bound 
emitters (either in molecular form or adsorbed on a substrate) suffers from an irreducible intrinsic energy uncertainty due to the emitter's zero-point motion. It was shown in~\cite{Cheipesh:2021fmg} that such an uncertainty is proportional to the dimensionless parameter $\gamma=[Q^2 m_{\text{e}}/
m_{\text{nucl}}^3]^{1/4}$, $Q$ being the energy released in the $\beta$ decay, $m_e, m_{\text{nucl}}$ - masses of the electron and nucleus respectively. It was also shown that this parameter is too large for $\isotope[3]{H},$ therefore Tritium-based detectors are unable to achieve the required energy resolution. Instead, the most promising candidates are $\Tm$ and $\Sm$ as they have the intrinsic energy uncertainty that is an order of magnitude lower than that of $\isotope[3]{H}$.

However, contrary to the case of $\isotope[3]{H}$ for which the neutrino cross section is known~\cite{Faessler:2011qj,Cocco:2007za}, theoretical calculation of $(\sigma v)_\nu$ for $\Tm$ and $\Sm$ poses a challenge. The quantum numbers (spin and parity) of the parent and daughter nuclei for these isotopes differ, hence the leptons are required to have a non-zero total orbital momentum. The latter can be composed in a non-unique way, which results in several different unknown nuclear constants entering the matrix element~\eqref{eq:matrix_element} that do not factor out.

We propose a way to estimate the relic neutrino capture cross section. Our proposal relies on the experimental measurement of the spectrum of $\beta$-decay near the endpoint. We show, that the extraction of the relic neutrino cross section can be achieved using the experimental data (via Eq.~\eqref{eq:final_estimate}) even if the energy resolution $\Delta E$ of the experiment that is much larger than neutrino mass $\Delta E \gg m_\nu$.

Finally, to get a rough idea of the feasibility of the relic neutrino capture experiment based on $\Tm$ ($Q = \unit[96.5]{keV}$,  $\tau = \unit[2.77]{years}$) or $\Sm$ ($Q = \unit[76.6]{keV}$,  $\tau = \unit[130]{years}$), we estimate the corresponding cross-sections using the $\beta$-decay spectra computed in BetaShape~\cite{Mougeot:2015bva, mougeot_be_bisch_2014}. For non-unique transitions, this software assumes that the $\xi$-approximation holds and evaluates the electromagnetic corrections to the spectrum.
\begin{align}\label{eq:cross_estimate}
    (\sigma v)_\nu\begin{cases} \unit[1.2 \times 10^{-46}]{cm^2} & \Tm\\ \unit[4.8 \times 10^{-48}]{cm^2}, & \Sm \end{cases} 
\end{align}

The corresponding neutrino capture rates per single atom $\Gamma_\nu = \eta_\nu (\sigma v)_\nu$ are:
\begin{align}\label{eq:bad_estimate}
    \frac{\Gamma_{\text{capture}}}{\unit{y^{-1}}} =\frac{\eta_\nu}{\la \eta_\nu\ra}\begin{cases}12.7\,(6.4) \times 10^{-27} & \Tm\\ 5.1\,(2.5) \times 10^{-28}, & \Sm \end{cases} 
\end{align}
for Majorana (Dirac) neutrino, where $\eta_\nu$ is the local cosmic number density of one neutrino species. This density could be significantly larger than the average over the universe $\langle \eta_\nu\rangle \sim \unit[56]{cm^{-3}}$ due to gravitational clustering. The corresponding cross-sections are in agreement with the crude estimate ($\delta \approx 0.5$).

Since the emitters in the solid-state based experiments are attached to the substrate atom by atom, the
single event exposure based on the estimate~\eqref{eq:bad_estimate} corresponds to $\unit[2\cdot 10^{27}]{atoms\cdot year}$ for $\Sm$ or $\unit[10^{26}]{atoms\cdot year}$ for $\Tm$. For comparison, the same number of events can be achieved with $\unit[2\cdot 10^{24}]{atoms\cdot year}$ for $\isotope[3]{H}$. According to this, using $\Tm$ as $\beta$ emitter in a full size C$\nu$B experiment is promising since it can provide with \textit{both} sufficient event rate \textit{and} energy resolution for the relic neutrino detection.

The estimates of the neutrino capture cross-section for $\Tm$ and $\Sm$ have been performed by two other theoretical groups after the completion of this work. In~\cite{Brdar:2022wuv}, the computation relies on the $\xi$-approximation accounting for possible deviations. In~\cite{Kostensalo:2022tcs}, the nuclear properties have been evaluated theoretically for $\Tm$, $\Sm$, and another potential candidate isotope $\isotope[210]{Pb}$. The latter was first analyzed in~\cite{deGroot:2022tbi} with the method, proposed in this work. The results of all studies agree with each other. However, contrary to our $O(1)$ precision, the results in~\cite{Brdar:2022wuv, Kostensalo:2022tcs} are potentially more precise, with the claimed uncertainties of the percent and ten percent level, respectively.

\begin{acknowledgements}
We would like to thank the theory group of PTOLEMY collaboration for many fruitful discussions. YC is supported by the funding from the Netherlands Organization for Scientific Research (NWO/OCW) and from the European Research Council (ERC) under the European Union’s Horizon 2020 research and innovation programme. AB is supported by the European Research Council (ERC) Advanced Grant “NuBSM” (694896). VC is grateful to the Dutch Research Council (NWO) for partial support, grant No 680-91-130. OM is supported by the funding from the NWO Physics Vrij Programme “The Hidden Universe of Weakly Interacting Particles” with project number 680.92.18.03 (NWO Vrije Programma), which is (partly) financed by the Dutch Research Council (NWO). 
\end{acknowledgements}

\bibliographystyle{spphys}       
\bibliography{bibliography}

\end{document}